%% file: main.tex
\documentclass[%
reprint,
showpacs,
showkeys,
nofootinbib,
nobibnotes,
amsmath,fixed
amssymb,
aps,
superscriptaddress
]{revtex4-1}

\usepackage[english]{babel}
\usepackage[utf8x]{inputenc}
\usepackage[T1]{fontenc}

\usepackage{amsmath}
\usepackage{graphicx}
\usepackage[export]{adjustbox}

\usepackage[colorlinks=true,linkcolor=blue,citecolor=blue,urlcolor=blue]{hyperref}

\usepackage{epstopdf}
\usepackage{amssymb}
\usepackage{color}
\usepackage{url}
\usepackage[justification=justified]{subcaption}
\usepackage[justification=justified, figurename=FIG.]{caption}

\usepackage{float}
\restylefloat*{figure}
\usepackage{placeins}

\usepackage{dcolumn}
\usepackage{bm} 

\usepackage{siunitx}
\usepackage{textcomp} 
\usepackage{multirow}
\usepackage{tabularx} 
\usepackage{ragged2e}
\usepackage[overload]{textcase}

\begin{document}

\title{New constraints on cosmic ray-boosted dark matter \\ from the LUX-ZEPLIN experiment}

\date{\today}

\input{authorlist}

\begin{abstract}
\noindent
While dual-phase xenon time projection chambers (TPCs) have driven the sensitivity towards weakly interacting massive particles (WIMPs) at the GeV/$c^2$ to TeV/$c^2$ mass scale, the scope for sub-GeV/$c^2$ dark matter particles is hindered by a limited nuclear recoil energy detection threshold.
One approach to probe for lighter candidates is to consider cases where they have been boosted by collisions with cosmic rays in the Milky Way, such that the additional kinetic energy lifts their induced signatures above the nominal threshold.
In this Letter, we report first results of a search for cosmic ray-boosted dark matter (CRDM) with a combined 4.2 tonne-year exposure from the LUX-ZEPLIN (LZ) experiment.
We observe no excess above the expected backgrounds and establish world-leading constraints on the spin-independent CRDM-nucleon cross section as small as $3.9 \times 10^{-33} \, \text{cm}^2$ at 90\% confidence level for sub-GeV/$c^2$ masses.
\end{abstract}


\maketitle
\enlargethispage{\baselineskip} 


Compelling astrophysical and cosmological evidence strongly supports the existence of dark matter (DM) in the Universe~\cite{Aghanim2020, Sofue:2000jx, Harvey:2015hha, Arbey:2021gdg}.
Despite numerous experimental efforts~\cite{XENON:2007uwm, ZEPLIN-III:2009htd, LUX:2016ggv, SuperCDMS:2017mbc, PICO:2017tgi, DEAP-3600:2017uua, XENON:2018voc, DarkSide:2018kuk, CDEX:2019hzn, DAMIC:2020cut, COSINE-100:2021zqh, XENON:2023cxc, SENSEI:2023zdf, CRESST:2024cpr, PandaX:2024qfu, LZ:2024lux} seeking to directly observe DM via scatters with nuclei, its detection has remained elusive~\cite{Bertone:2004pz, Roszkowski:2017nbc, Cirelli:2024ssz}.
Searches are complicated by a plethora of proposed candidates, spanning many orders of magnitude in mass.
For the favored weakly interacting massive particle (WIMP) hypothesis, dual-phase xenon time projection chambers (TPCs) have achieved unprecedented sensitivity for masses at the GeV/$c^2$-to-TeV/$c^2$ scale, down to DM-nucleon cross sections below $\sim \! 10^{-46}$~cm$^2$~\cite{XENON:2023cxc, PandaX:2024qfu, LZ:2024lux}.

As experimental constraints on WIMP-nucleon interactions approach the neutrino fog~\cite{Ohare:2021new}, attention has increasingly shifted towards exploring lower-mass DM candidates. 
However, as the kinetic energy of lighter DM particles---especially below the GeV/$c^2$ scale---becomes insufficient to produce detectable recoils on xenon nuclei, this parameter space remains less explored. 
Alternative detection channels have been considered in order to overcome this limitation, including ionization-only analyses~\cite{XENON:2019gfn, PandaX-II:2021nsg, XENON:2021qze, PandaX:2022xqx, XENON:2024znc}, the Migdal effect~\cite{LUX:2018akb, XENON:2019zpr, LZ:2023poo}, and inelastic scattering with associated photon emission~\cite{XENON10:2009sho, Aprile:2021inelastic}.

One intriguing avenue to access sub-GeV/$c^2$ DM involves leveraging boosted populations, where sufficient kinetic energy is imparted to generate detectable signals.
It has been posited that cold DM particles in the galactic halo could become relativistic through collisions with cosmic rays (CRs), producing a subpopulation of cosmic ray-boosted DM (CRDM)~\cite{Bringmann:2018cvk}.
In this description, the upscattering of DM involves the same DM-nucleus interaction mechanism as expected for direct detection experiments, thus requiring minimal model-dependent assumptions.
Various theoretical works have built upon this idea~\cite{Ema:2018bih, Cappiello:2019qsw, Dent:2019krz, Bondarenko:2019vrb, Ge:2020yuf, Xia:2020apm, Xia:2021vbz, Alvey:2022pad,Bardhan:2022bdg, Maity:2022exk, Xia:2022tid, Cappiello:2024acu, Dutta:2024kuj, Ghosh:2024dqw}, and experimental CRDM searches have been conducted or proposed with PROSPECT~\cite{PROSPECT:2021awi}, PandaX-II~\cite{PandaX-II:2021kai}, CDEX~\cite{CDEX:2022fig}, Super-Kamiokande~\cite{Super-Kamiokande:2022ncz} and NEWSdm~\cite{NEWSdm:2023qyb}, reporting constraints or sensitivities on spin-independent DM-nucleon contact interaction cross sections down to $\sim \! 10^{-32}$~cm$^2$, and thereby demonstrating the capability of terrestrial experiments to explore this new region of parameter space.

In this Letter, we utilize the combined 4.2 tonne-year exposure collected by the LUX-ZEPLIN (LZ) experiment thus far, as described in Ref.~\cite{LZ:2024lux}, to probe for interactions with sub-GeV/$c^2$ CRDM.
We incorporate the latest theoretical models for the upscattered CRDM flux {\color{black} as established in Refs.~\cite{Bringmann:2018cvk, Xia:2021vbz, Alvey:2022pad}} and comprehensive Monte Carlo simulations of Earth attenuation effects~\cite{Xia:2021vbz} to extend the sensitivity of LZ down by several orders of magnitude in mass.
These results serve to further solidify the position of LZ at the forefront of DM direct detection experiments.


The model of CRDM signatures in the LZ detector begins with a calculation of the CRDM flux at the surface of the Earth. 
To keep things generic, we follow Refs.~\cite{Bringmann:2018cvk} to consider a contact interaction between DM particles and nucleons with a constant cross section up to a form factor, which could arise from a heavy mediator.
We adopt the procedure established in Refs.~\cite{Bringmann:2018cvk, Xia:2021vbz} to obtain the differential CRDM flux at the surface, which can be expressed as
\begin{equation*}
    \frac{d\Phi^\text{loc}_\chi}{dT_\chi} = \frac{\rho^\text{loc}_\chi}{m_\chi}\sum_i F_i^2(Q^2)\sigma_{\chi i}\int^\infty_{T^\text{min}_i(T_\chi)} \frac{K_i(T_i)}{T_\chi^\text{max}(T_i)}\frac{d\Phi_i^\text{loc}}{dT_i} dT_i,
    \label{eq:surfaceFlux}
\end{equation*}
where $\rho^\text{loc}_\chi$ is the local DM density, $m_\chi$ is the DM particle mass, $T_\chi$ and $T_i$ denote the initial state kinetic energy of DM and CR particles respectively, $\sigma_{\chi i}$ is the spin-independent DM-nucleus scattering cross section, $F_i^2(Q^2)$ represents the nuclear form factor of CR nuclei as a function of the momentum transfer $Q^2$, and $i$ denotes different CR species.
The inhomogeneity of the primary CR distribution in the galaxy and the Navarro–Frenk–White DM profile~\cite{Navarro:1995iw} are accounted for by means of incorporating energy-dependent $K_i(T_i)$ factors as defined in Ref.~\cite{Xia:2021vbz}.
The differential local interstellar CR flux $d\Phi_i^\text{loc}/dT_i$ is adopted from the tabulated results in Refs.~\cite{Boschini:2020jty, Boschini:2017fxq}, originally obtained using the \textsc{GalProp-HelMod} framework~\cite{Moskalenko:1997gh, Strong:1998pw, Boschini:2017fxq}.
For the nuclear form factor $F_i$, following Refs.~\cite{Bringmann:2018cvk, Xia:2021vbz}, we assume the dipole form factor~\cite{Perdrisat:2006hj,Angeli:2004kvy} for hydrogen and helium, and the Helm form factor~\cite{Helm:1956zz,Lewin:1995rx} for heavier elements.
CR isotopes with atomic numbers from 3 (lithium) up to 28 (nickel) are included in the calculation of CRDM flux, which constitute approximately half of the total CRDM flux beyond just hydrogen and helium~\cite{Xia:2021vbz}.
It is also assumed that DM particles are point-like, such that a DM form factor is not required for the calculation.

Since the cross sections associated with CRDM interactions are at a significantly larger scale than those in conventional WIMP searches, the rock overburden of underground experiments introduces some attenuation to the CRDM flux~\cite{Bringmann:2018cvk, Xia:2021vbz, Alvey:2022pad}.
Previous studies have demonstrated that, while different treatments of attenuation effects generally yield similar lower bounds for the excluded region of parameter space, upper bounds can vary by several orders of magnitude depending on the assumed attenuation model~\cite{PandaX-II:2021kai, Alvey:2022pad, Super-Kamiokande:2022ncz}.

In this work, we employ the \textsc{DarkProp} Monte Carlo simulation framework outlined in Ref.~\cite{Xia:2021vbz} to model the attenuation of CRDM as it traverses the Earth's crust.
This approach enables stepwise simulation of the propagation, scattering, and angular deflection of DM particles within the Earth's crust, incorporating different nuclear form factors to ensure accurate propagation of attenuation effects. 
Here, the Earth is modeled as a homogeneous sphere that accounts for the chemical composition of the crust in evaluating the attenuated CRDM flux at the {\color{black} 4850~ft} depth of the LZ detector.

\begin{figure}[!t]
    \includegraphics[width=1.0\linewidth]{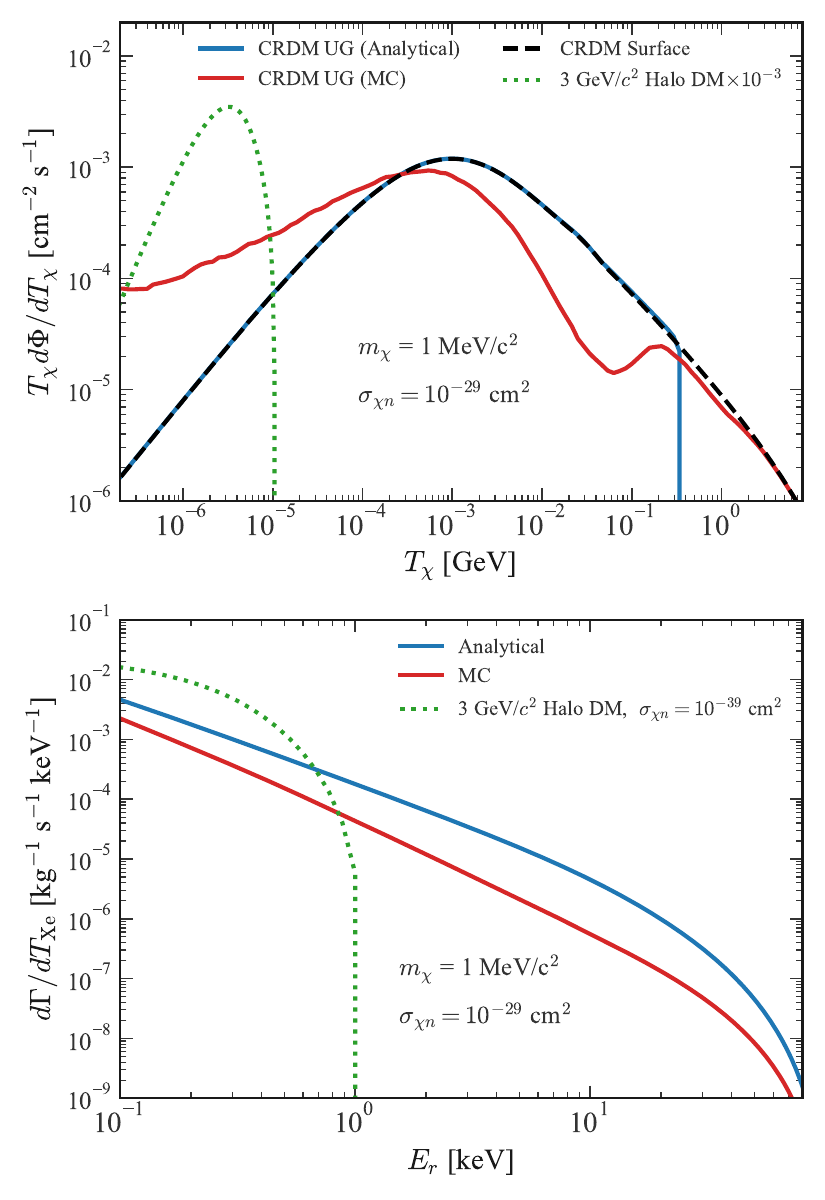}
    \caption{\justifying 
    The modeled CRDM flux for $m_\chi = 1$~MeV/c$^2$ and $\sigma_{\chi n} = 10^{-29}$~cm$^2$ (top) and corresponding nuclear recoil energy spectra in liquid xenon (bottom).
    The outcomes of two approaches for modeling the attenuated underground (UG) flux are shown for comparison: from Monte Carlo simulations (red), and with an analytical calculation using an energy loss method (blue).
    These are shifted towards lower energies with respect to the flux at the Earth's surface (dashed black).
    To illustrate the impact of boosting on the overall shape of each distribution, curves associated with a 3~GeV/$c^2$ WIMP (dotted green) are overlaid in both panels; the halo DM flux is scaled by a factor of $10^{-3}$ for visibility.
    For conventional WIMP searches, this mass is where sensitivity becomes limited as set by the recoil energy detection threshold~\cite{XENON:2024hup}.
    }
    \label{fig:fluxesAndEventRates}
\end{figure}

An alternative approach is the analytical energy loss-based model proposed in Ref.~\cite{Bringmann:2018cvk}, which omits the angular deflection and nuclear form factors for simplicity.
As nuclear form factors soften the attenuation effect and consequently lead to an overestimated underground CRDM flux, this simplification would generally lead to more conservative constraints~\cite{Xia:2021vbz,Alvey:2022pad}.
The surface and underground fluxes predicted by both attenuation models for $m_\chi = 1$~MeV/c$^2$ and a CRDM-nucleon cross section $\sigma_{\chi n} = 10^{-29}$~cm$^2$ are shown in the top panel of Figure~\ref{fig:fluxesAndEventRates}.
\textcolor{black}{In the analytical approach, the predicted underground flux exhibits a truncation at the high-energy tail, while the low-energy region remains largely unaffected. 
This feature arises from the fact that the energy loss rate per unit distance scales with $(T_\chi^2 + m_\chi T_\chi)$~\cite{Bringmann:2018cvk, Xia:2021vbz}. 
A more detailed discussion of the analytical attenuation model is available under Sec. III of Ref.~\cite{Xia:2021vbz}.
}
On the other hand, the Monte Carlo approach predicts an underground flux that is suppressed at mid-to-high energies compared to the surface, but enhanced at lower energies due to energy loss and downscattering of originally more energetic CRDM particles at the surface.
Higher cross sections lead to a more distorted underground CRDM flux in the Monte Carlo model, and a lower energy cut-off in the analytical model.
The impact of DM mass on the attenuation is more complicated: as the total cross section---and consequently the energy loss rate---is positively correlated with $m_\chi$~\cite{Xia:2021vbz} heavier CRDM particles are generally more attenuated by the overburden.
Nonetheless, results from both methods are in good agreement with calculations in Ref.~\cite{Xia:2021vbz}.

Following the formalism established in Ref.~\cite{Bringmann:2018cvk, Xia:2021vbz}, which assumes the same mechanism for CRDM scatters with xenon nuclei as in the initial boosting stage, the underground CRDM flux ${d\Phi^\text{UG}_\chi}/{dT_\chi}$ obtained from simulations at the depth of the LZ detector can be translated into a nuclear recoil energy spectrum according to
\begin{equation*}
    \frac{d\Gamma}{dT_\text{Xe}} = \mathcal{N}
    \int_{T_\chi^\text{min}(T_\text{Xe})}^\infty
    \sigma_{\chi n}
    \frac{F_\text{Xe}^2(Q^2)A_\text{Xe}^2}{T_\text{Xe}^\text{max}(T_\chi)}
    \frac{\mu^2_{\chi \text{Xe}}}{\mu^2_{\chi p}}
    \frac{d\Phi_\chi^\text{UG}}{dT_\chi}
    dT_\chi,
    \label{eq:eventRate}
\end{equation*}
where $\mathcal{N}$ is the number of target atoms per unit mass ($4.585\times10^{24}$ kg$^{-1}$ for xenon),  $\mu_{\chi \text{Xe}} = m_\chi m_\text{Xe} / (m_\chi + m_\text{Xe})$ denotes the reduced mass of a two-body elastic scatter between DM and a xenon nucleus of mass $m_\text{Xe}\sim122$~GeV/$c^2$.
The differential rate therefore depends on $m_\chi$ and $\sigma_{\chi n}$, which set the recoil spectra used as inputs for the signal simulations.
The bottom panel of Figure~\ref{fig:fluxesAndEventRates} illustrates this by displaying differential rates corresponding to the flux profiles shown in the top panel.
The LZ simulations chain~\cite{LZ:2021sim} samples the recoil spectra and utilizes NEST~\cite{Szydagis:2024nest}, as tuned to the LZ detector response with calibration data, to generate observables for the signal model.


The LZ experiment is situated 4850~ft underground within the Davis Cavern at the Sanford Underground Research Facility (SURF) in Lead, South Dakota, USA.
With a rock overburden equivalent to 4300~m of water, the experiment benefits from a factor of $3 \times 10^6$ reduction in the cosmic muon flux~\cite{Mei:2010early, Kudryavtsev:2009muon}, though this shielding now has the added relevance of CRDM flux attenuation.

As detailed in Refs.~\cite{LZ:2017tdr, LZ:2020nim, LZ:2022lsv, LZ:2024lux}, the LZ detector consists of a nested structure of both passive and active materials.
At its core, the detector consists of a cylindrical dual-phase xenon TPC with an active volume containing 7 tonnes of liquid xenon (LXe). 
Two anti-coincidence veto systems augment the detector: an instrumented 2-tonne LXe ``Skin'' surrounding the TPC is used to tag $\gamma$ rays, and an outer detector (OD) holding 17.3 tonnes of gadolinium-loaded liquid scintillator in a near-hermetic seal around the cryostat enables the rejection of neutron backgrounds~\cite{Haselschwardt:2019liquid}.
The entire apparatus is shielded from ambient radiation within a tank filled with 238 tonnes of ultra-pure water.

Energy depositions in the LXe target generate vacuum ultraviolet (VUV) prompt scintillation photons (S1) and ionization electrons.
An applied electric field drifts the electrons upwards, where they are extracted into a gaseous xenon phase by a stronger field and produce a delayed electroluminescence signal (S2).
Both signals are detected by arrays of photomultiplier tubes mounted at the top and bottom of the TPC.
The S2 hit pattern on the top array enables transverse $(x, y)$ position reconstruction, whereas the depth $(z)$ is informed by the drift time between the S1 and S2.
Moreover, the ratio of the two signals allows for discrimination between background-like electron recoils (ERs) and signal-like nuclear recoils (NRs).
Dispersed mono-energetic calibration sources such as $^{83\mathrm{m}}$Kr are deployed to normalize the detector response with respect to position, yielding corrected signals labeled as S1$c$ and S2$c$~\cite{LZ:2024calib}.
Furthermore, tritium $\beta$ decays and deuterium-deuterium (DD) neutrons are used to calibrate the ER and NR detector response, respectively.

\begin{figure}[!t]
    \includegraphics[width=1.0\linewidth]{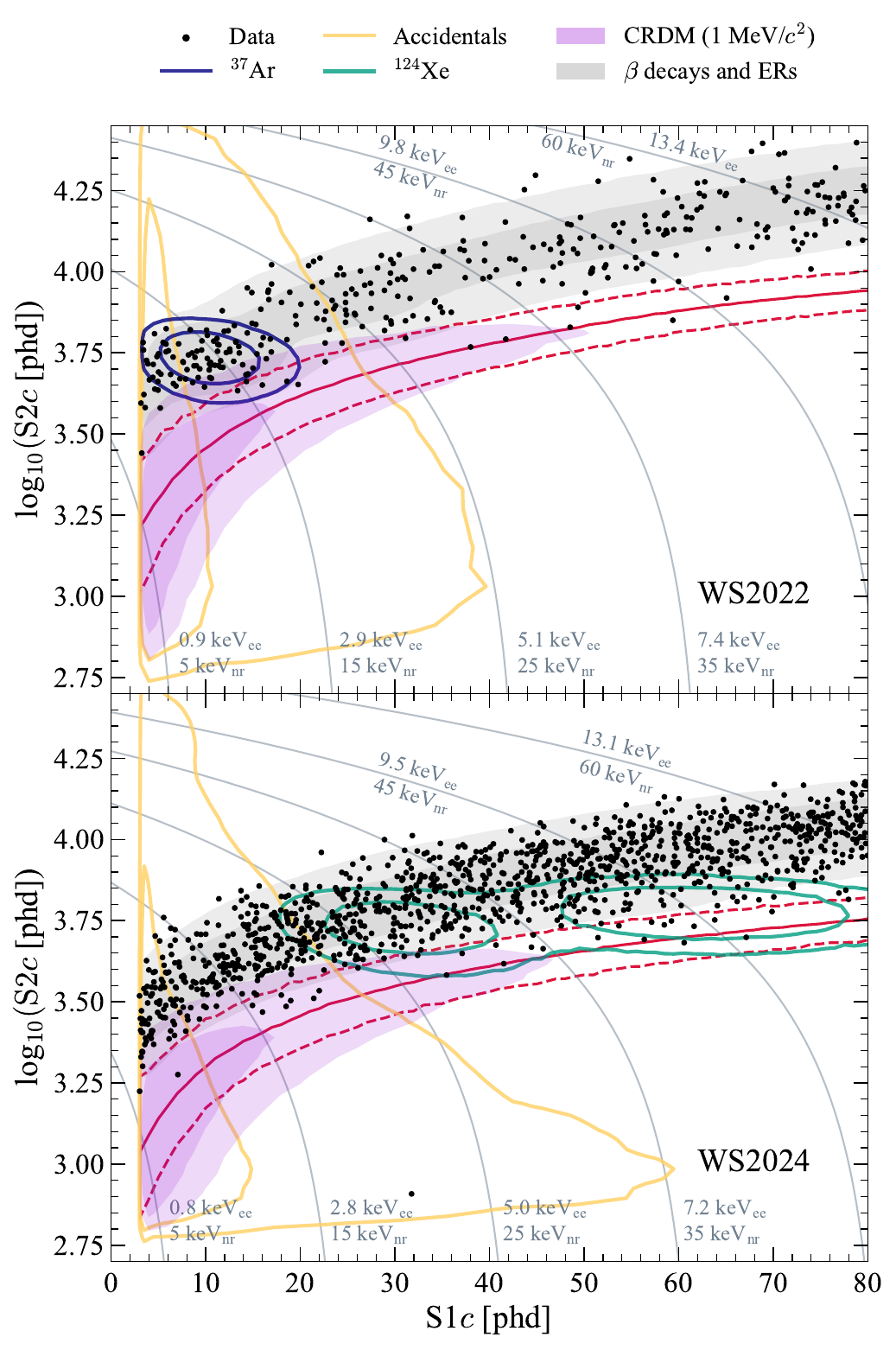}
    \caption{\justifying
    Final set of events (black points) passing all cuts for the 0.9 tonne-year WS2022 exposure (top) and the 3.3 tonne-year WS2024 exposure (bottom).
    Gray and purple shaded regions highlight the $1\sigma$ and $2\sigma$ contours for ER backgrounds and 1~MeV/$c^2$ CRDM, respectively.
    Contours are also drawn for distinct background sources: $^{37}$Ar from cosmogenic activation (navy), which is completely depleted by the start of WS2024; double electron captures of $^{124}$Xe (green) that dip towards the NR band due to enhanced recombination; and modeled accidental coincidence backgrounds (orange) in both runs.
    A red band marks the median NR response corresponding to the detector conditions in each dataset, along with 10\% and 90\% quantiles.
    Subtle differences can be seen for equivalent contours between WS2022 and WS2024, which reflect changes in the detector conditions and their subsequent modeling {\color{black}(akin to the supplementary material of Ref.~\cite{LZ:2024lux}), as well as the impact of changes in cut tunings.}}
    \label{fig:datapoints}
\end{figure}

For this analysis, we utilize the same final dataset from the combined 4.2 tonne-year LZ exposure as covered in Ref.~\cite{LZ:2024lux}, depicted in Figure~\ref{fig:datapoints}. 
These events were distilled from two separate runs: a 60 live-day exposure from the first LZ science run (WS2022), collected between December 2021 and May 2022, and a longer 220 live-day run (WS2024) spanning from March 2023 to April 2024.
The two campaigns are primarily distinguished by their differing detector conditions; the drift (extraction) field was lowered from 193~V/cm (7.3~kV/cm) for WS2022 to 97~V/cm (3.4~kV/cm) for WS2024, though with little overall impact on discrimination~\cite{Akerib2020:discrimination}.
Furthermore, a number of new features were successfully demonstrated in WS2024: a ``salting'' infrastructure to mitigate analyzer bias; a refined model of recombination enhancements in extremely rare double electron capture decays of $^{124}$Xe~\cite{LZ:2024dec, LZ:2025ec}; and a novel ``radon tag'' that targets $^{214}$Pb decays by means of tracking flow vectors of xenon as it circulates~\cite{LZ:2025radon}.


The statistical inference in this work follows an identical procedure to that in Ref.~\cite{LZ:2024lux}.
Fits to the data were performed with a two-sided unbinned profile likelihood ratio test statistic~\cite{Cowan:2011asymp}, conducted simultaneously on six mutually exclusive sub-samples.
One of these is the finalized WS2022 selection, unchanged from the first LZ result~\cite{LZ:2022lsv}, whereas the rest are attributed to the WS2024 exposure and describe events: (1) in a high-mixing circulation state; (2) in a low-mixing circulation state with an inactive radon tag; in a low-mixing circulation state that are either (3) radon tagged or (4) radon untagged; and (5) tagged by the Skin or OD vetos.
We adopt the same background model, such that the sole distinction is the choice of signal model, swapped from WIMPs to CRDM as per the treatment described previously.
{\color{black} As listed in Table \ref{tab:counts}, the best-fit counts are nearly identical to those obtained in the recent LZ WIMP search.
More details on the background model can be found in Refs.~\cite{LZ:2022lsv, LZ:SR1BG, LZ:2024lux}.}

\begin{table}[]
    \caption{\justifying \color{black} The pre-fit expectation and best-fit counts for all considered sources in the combined WS2022+WS2024 exposure, for a CRDM mass of 1 MeV$/c^2$. Although $\beta$-decay contributions were split up in WS2024 likelihood terms, they have been combined here for ease of comparability. $^{37}$Ar is not present in WS2024 as it will have been depleted since the 60 live day WS2022 campaign. Similarly, there is no atmospheric neutrino component under WS2022 due to the short exposure time.}
    \centering
    \begin{tabular}{p{5mm}lr@{}lr@{}l}
    \hline
    \hline
    \noalign{\vskip 1mm}
    \multicolumn{2}{c}{\textbf{Source}} & \multicolumn{2}{c}{\textbf{Expectation}} & \multicolumn{2}{c}{\textbf{Fit result}} \\
    \colrule \noalign{\vskip 1mm}
    \parbox[t]{2mm}{\multirow{8}{*}{\rotatebox[origin=c]{90}{WS2022}}} & $\beta$ decays + det. $\gamma$s & 215 &~$\pm$ 36 & 222 &~$\pm$ 16 \\
    & Solar $\nu$ ER & 27.1 &~$\pm$ 1.6 & 27.2 &~$\pm$ 1.6 \\
    & $^{136}$Xe $2\nu\beta\beta$ & 15.1 &~$\pm$ 2.4 & 15.2 &~$\pm$ 2.4 \\
    & $^{125}$Xe + $^{127}$Xe EC & 9.2 &~$\pm$ 0.8 & 9.3 &~$\pm$ 0.8 \\
    & $^{124}$Xe DEC & 5.0 &~$\pm$ 1.4 & 5.2 &~$\pm$ 1.4 \\
    & $^8$B $\nu$ NR & 0.14 &~$\pm$ 0.01 & 0.14 &~$\pm$ 0.01 \\
    & $^{37}$Ar & \multicolumn{2}{c}{$[0,\,288]$} & \multicolumn{2}{c}{$52.7^{+9.5}_{-8.9}$} \\
    & Accidental coincidences & 1.2 &~$\pm$ 0.3 & 1.2 &~$\pm$ 0.3 \\
    \colrule \noalign{\vskip 1mm}
    \parbox[t]{2mm}{\multirow{8}{*}{\rotatebox[origin=c]{90}{WS2024}}} & $\beta$ decays + det. $\gamma$s & 1026 &~$\pm$ 91 & 1017 &~$\pm$ 32 \\
    & Solar $\nu$ ER & 102 &~$\pm$ 6 & 102 &~$\pm$ 6 \\
    & $^{136}$Xe $2\nu\beta\beta$ & 55.6 &~$\pm$ 8.3 & 55.8 &~$\pm$ 8.2 \\
    & $^{125}$Xe + $^{127}$Xe EC & 3.2 &~$\pm$ 0.6 & 2.7 &~$\pm$ 0.6 \\
    & $^{124}$Xe DEC & 19.4 &~$\pm$ 3.9 & 21.4 &~$\pm$ 3.6 \\
    & $^8$B + \textit{hep} $\nu$ NR & 0.06 &~$\pm$ 0.01 & 0.06 &~$\pm$ 0.01 \\
    & Atm. $\nu$ NR & 0.12 &~$\pm$ 0.02 & 0.12 &~$\pm$ 0.02 \\
    & Accidental coincidences & 2.8 &~$\pm$ 0.6 & 2.6 &~$\pm$ 0.6 \\
    \colrule \noalign{\vskip 1mm}
    & Detector neutrons & \multicolumn{2}{c}{$0.0^{+0.2}$} & \multicolumn{2}{c}{$0.0^{+0.2}$} \\
    & 1 MeV$/c^2$ CRDM & \hfill &\phantom{0}-- & \multicolumn{2}{c}{$0.0^{+0.7}$} \\
    \colrule \noalign{\vskip 1mm}
    & Total & 1539 &~$\pm$ 107 & 1535 &~$\pm$ 36 \\
    \hline
    \hline
    \end{tabular}
    \label{tab:counts}
\end{table}

\begin{figure}[t]
    \includegraphics[width=1.0\linewidth]{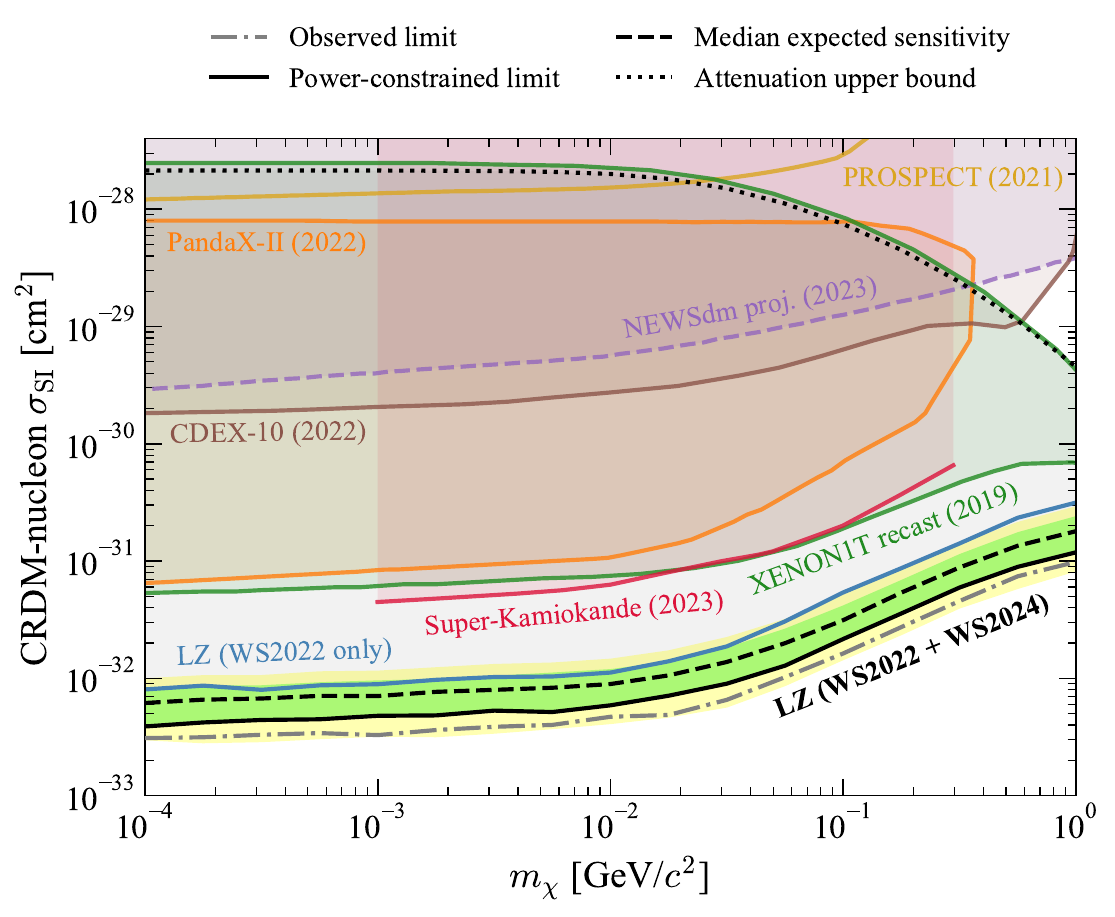}
    \caption{\justifying Observed upper limit on the spin-independent CRDM-nucleon cross section at 90\% confidence level as a function of CRDM mass from the combined 280 live-day WS2022+WS2024 exposure.
    The limit prior to the application of a power constraint is shown as a gray dot-dashed line.
    The median expected sensitivity for background-only experiments is drawn with a dashed black line, with corresponding $1\sigma$ and $2\sigma$ levels shaded as green and yellow bands, respectively.
    A dotted black line marks the upper bound on the cross section, derived analytically based on expectations of the CRDM flux attenuation.
    It should be emphasized that the region between this analytical upper bound and the observed upper limit forms the excluded parameter space.
    Along with the WS2022-only limit, relevant upper limits and excluded regions are also shown from PandaX-II~\cite{PandaX-II:2021kai}, CDEX-10~\cite{CDEX:2022fig}, PROSPECT~\cite{PROSPECT:2021awi}, and Super-Kamiokande~\cite{Super-Kamiokande:2022ncz}, as well as a recast for XENON1T~\cite{Bringmann:2018cvk} and projected sensitivity for NEWSdm~\cite{NEWSdm:2023qyb}.
    }
    \label{fig:limits}
\end{figure}

Figure~\ref{fig:limits} presents the observed 90\% confidence level upper limit on the spin-independent CRDM-nucleon cross section as a function of mass, contextualized with recent experimental limits and sensitivity projections.
Following the conventions set by the community in Ref.~\cite{Baxter:2021rec}, the limit is power constrained to $1\sigma$ below the median at all masses considered.
This is due to the fact that all CRDM spectra overlap with background sources that are noted to have underfluctuated: $^{37}$Ar in WS2022~\cite{LZ:2022lsv}, and accidental coincidences in WS2024.
Nevertheless, the fitted nuisance parameters obtained here are in excellent agreement with those obtained in the WIMP search analysis.
The best-fit number of CRDM events for all masses tested between 100~keV/$c^2$ and 1~GeV/$c^2$ is zero.
The data are thus consistent with the background-only hypothesis, with the goodness of fit assessed across a range of metrics such as reconstructed energy and distance to the ER band median, as well as in $\left\{\text{S1}c, \log_{10}(\text{S2}c) \right\}$ space.
The model-data agreement is further verified using a Holm-Bonferroni test~\cite{Holm:1979test}, with all samples passing at a significance level of 0.05.

The range of CRDM masses evaluated here spans between 100~keV/$c^2$ and 1~GeV/$c^2$.
In principle, lighter masses could be considered~\cite{Xia:2020apm}, though their existence would present additional implications.
For one, it has been argued that the abundance of low-mass DM would have influenced the production of light elements in the early Universe, except this is model-dependent and only applies when Standard Model decay modes exist~\cite{Alvey:2022pad, PandaX-II:2021kai}.
Another constraint stems from the Pauli exclusion principle, requiring that fermionic DM cannot be lighter than 0.1~keV/$c^2$~\cite{DiPaolo:2017geq}.
On the other hand, truncating the CRDM mass at 1~GeV/$c^2$ reflects the degree of certainty associated with the signal model.
In particular, the total DM-nucleus cross section $\sigma_{\chi N}$ saturates at the geometric cross section of the nucleus $4\pi r_A^2$, where $r_A$ is the radius of the nucleus, for higher CRDM masses~\cite{Digman:2019wdm, Dutta:2024kuj}.
In turn, this invalidates the $A^2\mu^2$ coherent enhancement to the cross section for $\sigma_{\chi n} > 4 \times 10^{-28}$~cm$^2$ and $m_\chi > 1$~GeV$/c^2$ in the context of xenon-based experiments~\cite{Digman:2019wdm, Bell:2023sdq}.

Throughout this work, it is assumed that the DM-nucleon interaction cross section is energy independent up to a form factor, which can be described with a contact interaction or a heavy mediator in a low momentum transfer limit~\cite{Digman:2019wdm, Alvey:2022pad}.
This leads to more conservative constraints than energy dependent approaches, especially for smaller DM masses~\cite{Dent:2019krz}.
Alternatively, large total cross sections above the geometric size limit can be realized through light mediators, though their interactions with Standard Model fermions are subject to stringent but model-dependent limits from monojet searches, meson decays, and stellar cooling~\cite{Alvey:2022pad, Bell:2023sdq, Cappiello:2024acu}, with vector mediators typically facing stronger constraints than scalars~\cite{Cappiello:2019qsw, Ema:2020ulo}.
These constraints primarily arise from couplings to quarks, while alternative interaction portals, including Higgs and gluons, may lead to different phenomenological constraints~\cite{Xu:2024iny, Elor:2021swj}.

We report an improvement in the upper limit on the CRDM-nucleon cross section by over one order of magnitude relative to recent limits~\cite{PandaX-II:2021kai, CDEX:2022fig, PROSPECT:2021awi, Super-Kamiokande:2022ncz}.
This can be ascribed to a larger accumulated exposure, significantly lower background rates, the implementation of novel techniques such as the radon tag, and the treatment of contributions from heavier CR species beyond hydrogen and helium.
In recent studies~\cite{PROSPECT:2021awi, PandaX-II:2021kai, CDEX:2022fig, Alvey:2022pad, Bringmann:2018cvk}, an explicit upper bound on the excluded parameter space is often included. 
However, this is heavily dependent on the attenuation model, as covered in recent theoretical and experimental works~\cite{PandaX-II:2021kai, CDEX:2022fig, Alvey:2022pad}.
We also note that the distribution of the primary CR itself begins to change at $\sigma_{\chi n} > 10^{-27}$~cm$^2$, causing high-rigidity CR spectra to deviate from experimental data ~\cite{Ge:2020yuf, Cappiello:2018hsu}.
Moreover, CRDM models with such large $\sigma_{\chi n}$ can saturate the total DM-nucleus cross section~\cite{Digman:2019wdm}, while only being moderately attenuated in the overburden, as per our simulations and Ref.~\cite{Super-Kamiokande:2022ncz}.
As a result, we follow Ref.~\cite{Super-Kamiokande:2022ncz} in refraining from quoting an official upper bound from a statistical treatment. Instead, we provide a reference analytical bound in Figure~\ref{fig:limits}, which is calculated from the analytical energy loss method formulated in Refs.~\cite{Bringmann:2018cvk, Xia:2021vbz} while ignoring the nuclear form factor of elements in the Earth's crust.
This produces a more conservative upper bound than any other treatment~\cite{Xia:2021vbz, Alvey:2022pad}.


In conclusion, we present first results from a search for CRDM in LZ with the accumulated 4.2 tonne-year exposure employed in the recent WIMP search analysis~\cite{LZ:2024lux}.
No significant excess over the expected background levels is observed, and a world-leading limit is set on spin-independent CRDM-nucleon cross sections over a range of sub-GeV/$c^2$ masses.

\textit{Acknowledgements} -- The authors thank Yonglin Li and Zuowei Liu for fruitful discussions. The research supporting this work took place in part at the Sanford Underground Research Facility (SURF) in Lead, South Dakota. Funding for this work is supported by the U.S. Department of Energy, Office of Science, Office of High Energy Physics under Contract Numbers DE-AC02-05CH11231, DE-SC0020216, DE-SC0012704, DE-SC0010010, DE-AC02-07CH11359, DE-SC0015910, DE-SC0014223, DE-SC0010813, DE-SC0009999, DE-NA0003180, DE-SC0011702, DE-SC0010072, DE-SC0006605, DE-SC0008475, DE-SC0019193, DE-FG02-10ER46709, UW PRJ82AJ, DE-SC0013542, DE-AC02-76SF00515, DE-SC0018982, DE-SC0019066, DE-SC0015535, DE-SC0019319, DE-SC0024225, DE-SC0024114, DE-AC52-07NA27344, \& DE-SC0012447. This research was also supported by U.S. National Science Foundation (NSF); the UKRI’s Science \& Technology Facilities Council under award numbers ST/W000490/1, ST/W000482/1, ST/W000636/1, ST/W000466/1, ST/W000628/1, ST/W000555/1, ST/W000547/1, ST/W00058X/1, ST/X508263/1, ST/V506862/1, ST/X508561/1, ST/V507040/1 , ST/W507787/1, ST/R003181/1, ST/R003181/2,  ST/W507957/1, ST/X005984/1, ST/X006050/1; Portuguese Foundation for Science and Technology (FCT) under award numbers PTDC/FIS-PAR/2831/2020; the Institute for Basic Science, Korea (budget number IBS-R016-D1); the Swiss National Science Foundation (SNSF)  under award number 10001549. This research was supported by the Australian Government through the Australian Research Council Centre of Excellence for Dark Matter Particle Physics under award number CE200100008. We acknowledge additional support from the UK Science \& Technology Facilities Council (STFC) for PhD studentships and the STFC Boulby Underground Laboratory in the U.K., the GridPP~\cite{Faulkner:2005gridpp, Britton:2009gridpp} and IRIS Collaborations, in particular at Imperial College London and additional support by the University College London (UCL) Cosmoparticle Initiative, and the University of Zurich. We acknowledge additional support from the Center for the Fundamental Physics of the Universe, Brown University. K.T. Lesko acknowledges the support of Brasenose College and Oxford University. The LZ Collaboration acknowledges the key contributions of Dr. Sidney Cahn, Yale University, in the production of calibration sources. This research used resources of the National Energy Research Scientific Computing Center, a DOE Office of Science User Facility supported by the Office of Science of the U.S. Department of Energy under Contract No. DE-AC02-05CH11231. We gratefully acknowledge support from GitLab through its GitLab for Education Program. The University of Edinburgh is a charitable body, registered in Scotland, with the registration number SC005336. The assistance of SURF and its personnel in providing physical access and general logistical and technical support is acknowledged. We acknowledge the South Dakota Governor's office, the South Dakota Community Foundation, the South Dakota State University Foundation, and the University of South Dakota Foundation for use of xenon. We also acknowledge the University of Alabama for providing xenon. For the purpose of open access, the authors have applied a Creative Commons Attribution (CC BY) license to any Author Accepted Manuscript version arising from this submission. Finally, we respectfully acknowledge that we are on the traditional land of Indigenous American peoples and honor their rich cultural heritage and enduring contributions. Their deep connection to this land and their resilience and wisdom continue to inspire and enrich our community. We commit to learning from and supporting their effort as original stewards of this land and to preserve their cultures and rights for a more inclusive and sustainable future.

\textit{Appendix--Data availability} Selected data from this analysis are publicly available \cite{hepdata.155182, hepdata.157863}, including the following:
\begin{itemize}
    \item Figure.~\ref{fig:datapoints}: all the events passing data analysis cuts in the WS2022+WS2024 combined dataset in $\left\{\text{S1}c, \log_{10}(\text{S2}c) \right\}$.
    \item Figure.~\ref{fig:limits}: The observed upper limit and sensitivity bands as a function of $m_\chi$.
\end{itemize}

\bibliographystyle{apsrev4-2}
\bibliography{references}

\end{document}

%% file: authorlist.tex

\author{J.~Aalbers}
\affiliation{SLAC National Accelerator Laboratory, Menlo Park, CA 94025-7015, USA}
\affiliation{Kavli Institute for Particle Astrophysics and Cosmology, Stanford University, Stanford, CA  94305-4085 USA}

\author{D.S.~Akerib}
\affiliation{SLAC National Accelerator Laboratory, Menlo Park, CA 94025-7015, USA}
\affiliation{Kavli Institute for Particle Astrophysics and Cosmology, Stanford University, Stanford, CA  94305-4085 USA}

\author{A.K.~Al Musalhi}
\email{aiham.almusalhi@ucl.ac.uk}
\affiliation{University College London (UCL), Department of Physics and Astronomy, London WC1E 6BT, UK}

\author{F.~Alder}
\affiliation{University College London (UCL), Department of Physics and Astronomy, London WC1E 6BT, UK}

\author{C.S.~Amarasinghe}
\affiliation{University of California, Santa Barbara, Department of Physics, Santa Barbara, CA 93106-9530, USA}

\author{A.~Ames}
\affiliation{SLAC National Accelerator Laboratory, Menlo Park, CA 94025-7015, USA}
\affiliation{Kavli Institute for Particle Astrophysics and Cosmology, Stanford University, Stanford, CA  94305-4085 USA}

\author{T.J.~Anderson}
\affiliation{SLAC National Accelerator Laboratory, Menlo Park, CA 94025-7015, USA}
\affiliation{Kavli Institute for Particle Astrophysics and Cosmology, Stanford University, Stanford, CA  94305-4085 USA}

\author{N.~Angelides}
\affiliation{Imperial College London, Physics Department, Blackett Laboratory, London SW7 2AZ, UK}

\author{H.M.~Ara\'{u}jo}
\affiliation{Imperial College London, Physics Department, Blackett Laboratory, London SW7 2AZ, UK}

\author{J.E.~Armstrong}
\affiliation{University of Maryland, Department of Physics, College Park, MD 20742-4111, USA}

% ...

\collaboration{The LZ Collaboration}